\documentclass[useAMS,usenatbib]{article} 
\usepackage{epsf}
\usepackage{pazh}
\usepackage{graphicx}
\usepackage{wrapfig}
\usepackage{natbib}
\usepackage{hyperref}
\usepackage{gensymb}
\def\grs{{GRS\,1739-278\,}}

\tightenlines
\sloppypar

\begin{document}

\title{\bf The outburst of the X-ray nova GRS\,1739-278 in September, 2016}

\author{\bf \hspace{-1.3cm}\ \ 
I.A. Mereminskiy\affilmark{1}, E.V. Filippova\affilmark{1}, R.A. Krivonos\affilmark{1}, S.A. Grebenev\affilmark{1}, R.A. Burenin\affilmark{1} and R.A.~Sunyaev\affilmark{1,2}}
$^{1}$Space Research Institute, Russian Academy of Sciences, Profsoyuznaya 84/32, 117997 Moscow, Russia\\
$^{2}$Max Planck Institute for Astrophysics, Karl-Schwarzschild-Strasse 1, D-85741 Garching, Germany\vspace{2mm}

\sloppypar 
\vspace{2mm}
\noindent
During the scanning observations of the Galactic Center region in August -- September 2016 we detected the new outburst of the historical X-ray nova \grs, the black hole candidate LMXB system. In this letter we present results of INTEGRAL and Swift-XRT observations taken during the outburst. In hard X-ray band (20--60 keV) the flux from the source raised from $\sim$11 to $\sim$30 mCrab between 3 and 14 of September. For nearly 8 days the source has been observed at this flux level and then faded to $\sim$15 mCrab. The broadband quasi-simultaneous spectrum obtained during the outburst is well described by the absorbed powerlaw with the photon index $\Gamma=1.86\pm0.07$ in broad energy range 0.5--150 keV,  with absorption corresponding to ${N_{H}}=2.3\times10^{22}$~cm$^{-2}$ assuming solar abundance. Based on this we can conclude that the source was in the low/hard state. From the lightcurve and spectra we propose that this outburst was `failed', i.e. amount of accreted matter was not sufficient to achieve the high/soft spectral state with dominant soft blackbody component as seen in normal outbursts of black hole candidates. 

\\[0.3cm]

Accepted for publication in {\em Astronomy Letters, 2017, n.3}.

\vfill
\noindent\rule{8cm}{1pt}\\
{$^*$Corresponding author $<$i.a.mereminskiy@gmail.com$>$}

\clearpage

\section{Introduction}
\label{sec:intro} 

The X-ray nova \grs was discovered by SIGMA telescope onboard GRANAT space observatory during its bright outburst in 1996 \citep{paul96}. Later GRANAT continued observations of this source \citep{vargas97} along with other X-ray telescopes: ROSAT \citep{greiner96}, RXTE and TTM/Kvant \citep{borozdin98}. The optical counterpart was found in observations carried out by ESO telescopes \citep{marti97}. The peak flux was about $\simeq800$ mCrab in 2--10 keV (ASM/RXTE) \citep{borozdin98}. During the outburst \grs demonstrated typical behavior of the black hole candidate \citep[BHC, see e.g.][]{grebenev93, grebenev97, tanaka96, remillard06, belloni10}. The lightcurve could be described with a FRED-like (fast rise, exponential decay) shape, at the beginning of the outburst the source was in a typical low/hard state, with spectrum dominated by a power-law component with exponential cutoff at high ($\ge$100 keV) energies followed by a high/soft state with a strong blackbody component \citep{borozdin98}. Observations at VLA \citep{durouchoux96} revealed presence of variable radio emission, which could be caused by jets.  \cite{borozdin00} found QPO at 5 Hz in RXTE observations conducted while the source was in the very high/soft state.

Since the discovery of the source in 1996, the Galactic Center region has been regularly monitored by RXTE/ASM and INTEGRAL (since 2003) missions. According to \cite{krivonos12} \grs remained in quiescence until 2013, the upper limit on 20--60 keV flux was 0.12 mCrab (3$\sigma$, 1 mCrab corresponds to 12.3$\times$10$^{-12}$ erg cm$^{-2}$ s$^{-1}$). 

\begin{figure}
\centerline{\includegraphics[scale=0.8]{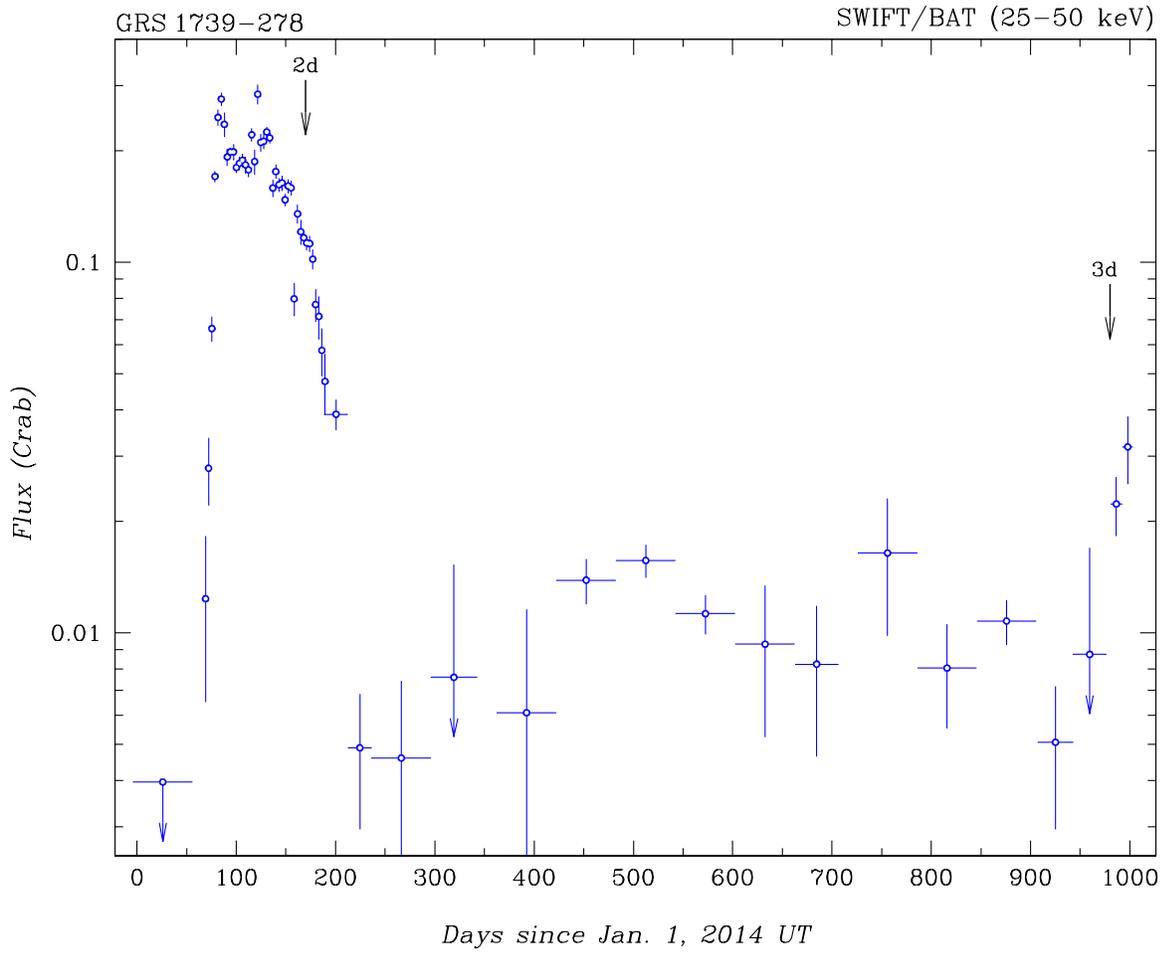}}
\caption{Swift-BAT lightcurve in 25--50 keV range taken between January, 2014 and September, 2016. The second large outburst (denoted as `2d') and onset of the third outburst (denoted as `3d') are clearly visible. During outbursts time bin was choosen as 6 days and between outbursts as 2 months.} 
\label{fig:batlc}
\end{figure}

The second outburst of \grs was detected by \cite{filippova14} in 2014. It's 2--20 keV flux was measured at the level of $\sim$200 mCrab\footnote{http://maxi.riken.jp/top/index.php?cid=1\&jname=J1742-277} with MAXI experiment \citep{matsuoka09}. Fig.~\ref{fig:batlc} shows the source lightcurve during the second outburst as measured by Swift-BAT telescope \citep{gehrels04bat} in 25-50 keV. One can notice that although the second outburst apparently ended after $\sim$150 days after the beginning, the source flux has not lowered down to zero and remained on 5--15 mCrab level. As shown in Fig.~\ref{fig:batlc} in the beginning of September the flux from \grs started to increase. This allowed us to report an onset of the new outburst using INTEGRAL observations \citep{mereminskiy16}. In this work we used all publicly available INTEGRAL data, including monitoring observations of the Galactic Bulge \citep{kuulkers07_gbm} and private data from scanning observations of the Galactic Center \citep[as described in ][]{krivonos12antic}. We also used two Swift-XRT observations performed on September 21 and 24 as well as follow-up observations by Russian-Turkish 1.5-m telescope (RTT-150) on September, 26.

\section{Observations}
\begin{figure}
\centerline{\includegraphics[scale=0.95]{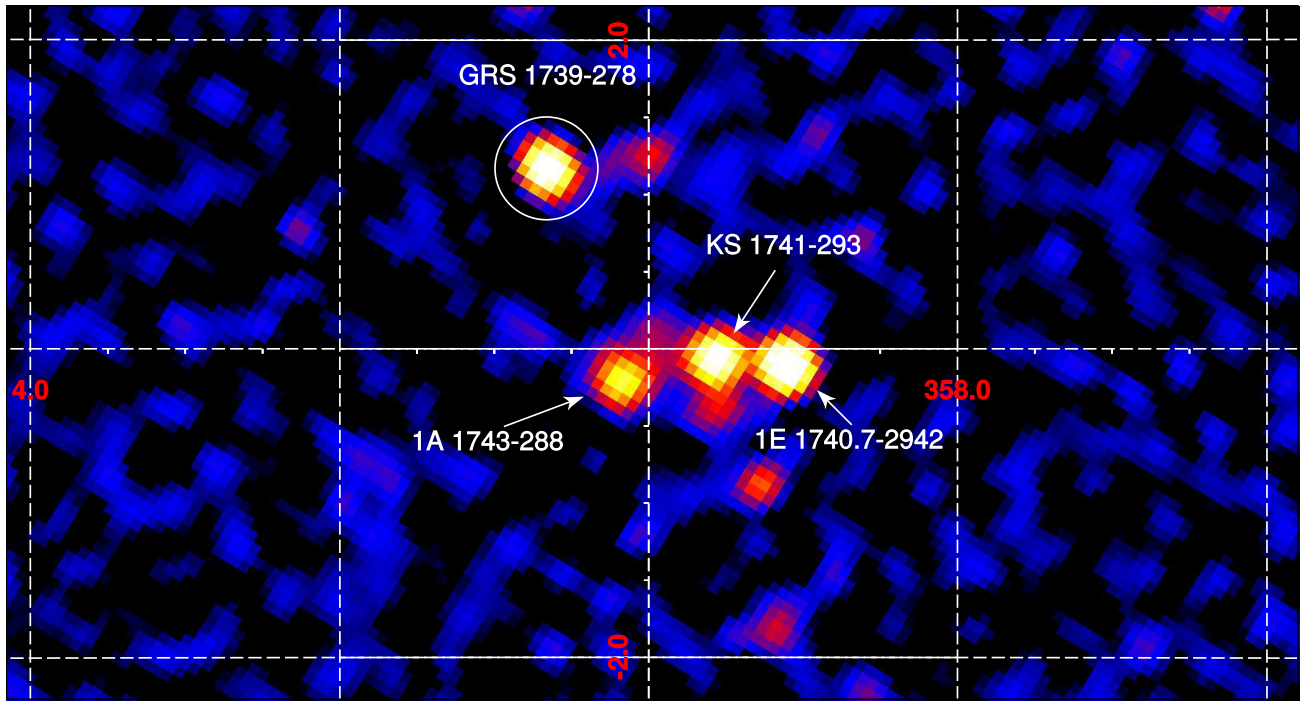}}
\caption{The image of the Galactic Center region in 20-60 keV band obtained with IBIS/ISGRI (Galactic coordinates, shown in term of significance). The square-root color map ranges from 0 to 25. The mosaic was made using observations conducted from 14 to 21 September 2016.} 
\label{fig:ima}
\end{figure}

Main results were obtained with IBIS/ISGRI telescope \citep{ubertini03} onboard INTEGRAL observatory \citep{winkler03}. IBIS/ISGRI is a wide-field ($30\degree\times30\degree$ FWZR) coded mask telescope working in the hard X-ray range 20--300 keV. The angular resolution of about 13' FWHM allows a confident detection of \grs despite its location in the Galactic Bulge, a region crowded by a large number of bright point X-ray sources (\cite{krivonos12}, see also Fig.~\ref{fig:ima} for illustration). We also used JEM-X telescope onboard INTEGRAL, which is sensitive in the standard X-ray range 3--35 keV  \citep{lund03} and has field of view of 13.2$\degree$ in diameter.
The field around the source was observed by INTEGRAL from 29 August until 27 September 2016 which corresponds to 1719-1729 INTEGRAL orbits, with exception for 1723 and 1724 revolutions, when INTEGRAL observed Crab nebulae for calibration purposes. 

For IBIS/ISGRI data we performed energy calibration (processing of event lists up to the COR level) with use of OSA 10.1 \citep{courvoisier2003}. Then we used the proprietary analysis package developed at IKI  \citep{revnivtsev04,krivonos10,churazov14} to reconstruct sky images and extract source fluxes. JEM-X  data were reduced with the standard software OSA 10.1 to obtain sky images (IMA level) and then processed according to \cite{grebenev15}.  

Swift-XRT observational campain was initiated immediately \citep{neilsen16} after discovery of the new outburst. The first observation (ObsID.~00033812055, hereafter we use only two last digits of the ObsID.) was performed on 21 September (57652 MJD), XRT observed \grs in Photon Counting mode for $\sim$1 ks. The second observation (ID.~56) took place on 24 September (57656 MJD). We processed all data using the standard pipeline XRTPIPELINE v.0.12.6 \citep{burrows05}. The exposure map was build by XRTEXPOMAP and was used to produce ARF files. XRT observed \grs at nearly 1.5 cts s$^{-1}$ (0.5--10 keV) rate. For the spectral analysis we had rebinned data in order to have at least 100 photons per bin. Unfortunately, the source count rate was too low to study its variability at timescales of minutes or seconds.

On 26 September (16:50 UTC) we performed a search for an optical/IR emission with RTT-150 telescope with the focal reducer and \emph{TFOSC} spectrograph. Observational conditions were poor because of the large zenith angle, the image quality was about 2.5$^{\prime\prime}$. We got direct images in $r^\prime$ and $i^\prime$ band , the exposure time was 300 s.

\section{Results}

Fig.~\ref{fig:ima} shows the significance map of the field around \grs obtained with IBIS/ISGRI in 20--60 keV range on September 2016 (spacecraft revolutions 1725-1727). The source is detected at 34$\sigma$ significance level, the exposure on the source is 62 ksec. Several other bright X-ray sources in the Galactic Bulge are also marked.

\begin{figure}
\centerline{\includegraphics[scale=0.7]{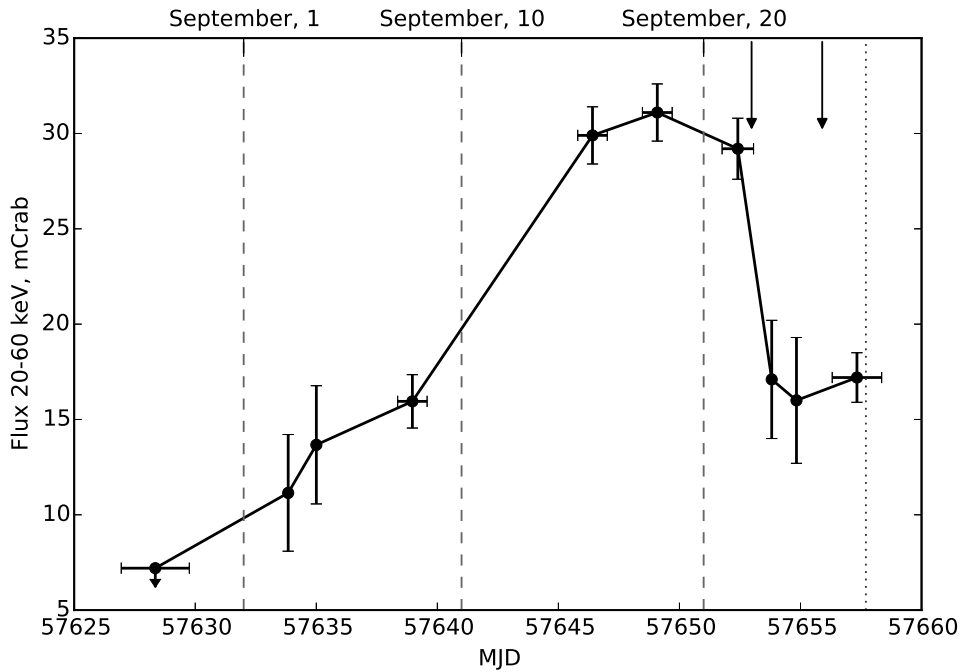}}
\caption{The hard X-ray lightcurve (20--60 keV) of \grs during the third outburst as seen by INTEGRAL. Arrows show coordinated Swift-XRT observations, the dotted line represents time of our optical/IR follow-up.} 
\label{fig:lc}
\end{figure}

In Fig.~\ref{fig:lc} we present the lightcurve of the source in 20--60 keV band taken with IBIS/ISGRI for all available data. The first significant detection of the source occured on 3 September, the flux was $11.1\pm3.1$ mCrab. Several features are clearly seen on the figure: the slow and steady increase of the flux for about 10 days, the ``plateau'' at $\sim30$ mCrab level that lasted for a week, and, then, the decline to $\sim$15 mCrab. The source remained at this stage until the end of our observations. This behaviour is completely different from that in the previous outburst of the source, during which intensity in 15--50 keV band increased tenfold in the course of 8 days and reached $\sim$300 mCrab at maximum as Swift-BAT data show \citep{krimm14_atel} (see also Fig.~\ref{fig:batlc}).

We reconstructed and studied spectra obtained by IBIS/ISGRI in 1725, 1726 and 1727 orbits, during which the source was bright. We also built the overall spectrum for 1728-1729 revolutions. To fit spectra we used {\em wabs*powerlaw} model from \texttt{XSPEC} package \citep{arnaud96}. All spectra are well-fitted by the powerlaw with the slope of $\,\sim$1.7--2.0 (Table~\ref{tab:perrev}), no high-energy cut-off is seen up to 150 keV.

\begin{table*}
\caption{The best-fit spectral parameters for IBIS/ISGRI observations of \grs}              
\label{tab:perrev}       
\centering  
\scriptsize                                    
\begin{tabular}{@{}c c c c c c @{}}    
\hline\hline                        
Revolution & Obs. start    & Obs. end &Exposure, & Flux 20-100 keV,                                      & {\bf $\Gamma$},\\    
                  &MJD              & MJD        &ks & 10$^{-10}$ egr cm$^{-2}$ s$^{-1}$   & photon index                          \\ 
\hline
1725              &   57645.8 & 57647.0& 21 & 5.8$\pm$2.6  & 1.73$\pm$0.12\\
1726               &   57648.5 & 57649.7&23 & 5.6$\pm$2.6 & 1.98$\pm$0.12\\
1727               &  57651.9 & 57653.1 &19 & 5.2$\pm$2.8 & 1.84$\pm$0.14\\
1728-1729        & 57653.7  & 57658.3& 37 & 3.0$\pm$1.9  & 1.81$\pm$0.16\\
\hline                                             
\end{tabular}
\vspace{-0.2truecm}
\end{table*}
 
In Swift-XRT observations the source was already in fading phase, yet still confidiently detected at 1.28$\pm$0.04 cts s$^{-1}$ ({\it S/N} $\simeq$ 30). Spectra obtained by Swift-XRT show no obvious peculiarities and can be described with the absorbed powerlaw. The measured absorption column $\sim$2.3$\times$10$^{22}$ cm$^{-2}$ is in agreement with previous estimates \citep{greiner96,miller15_nust}. The best-fit models are presented in Table~\ref{tab:swift}. The slope of the powerlaw is measured with large errors due to narrow width of the energy range and strong correlation with the absorption.

\begin{table*}
\caption{The best-fit spectral parameters of \grs from Swift-XRT spectra}              
\label{tab:swift}       
\centering  
\scriptsize                                    
\begin{tabular}{@{}c c c c c c@{}}    
\hline\hline                        
ID & Obs. start & Exposure,& Unabsorbed flux (0.1-10 keV),                                                        & {\bf $\Gamma$}, & N$_{H},$\\    
             &MJD                   & s & 10$^{-10}$, erg cm$^{-2}$ s$^{-1}$   &        photon index                    & 10$^{22}$ cm$^{-2}$\\ 
\hline
55 &   57652 &968.9& 2.4$\pm$0.6 & 1.74$\pm$0.24 & 2.14$\pm$0.43\\
56 &   57656 &939& 1.6$\pm$0.2  & 1.5$\pm$0.3 & 1.52$\pm$0.19\\
\hline                                             
\end{tabular}
\vspace{-0.2truecm}
\end{table*}

\begin{figure}[h]
\centering
\includegraphics[scale=0.7]{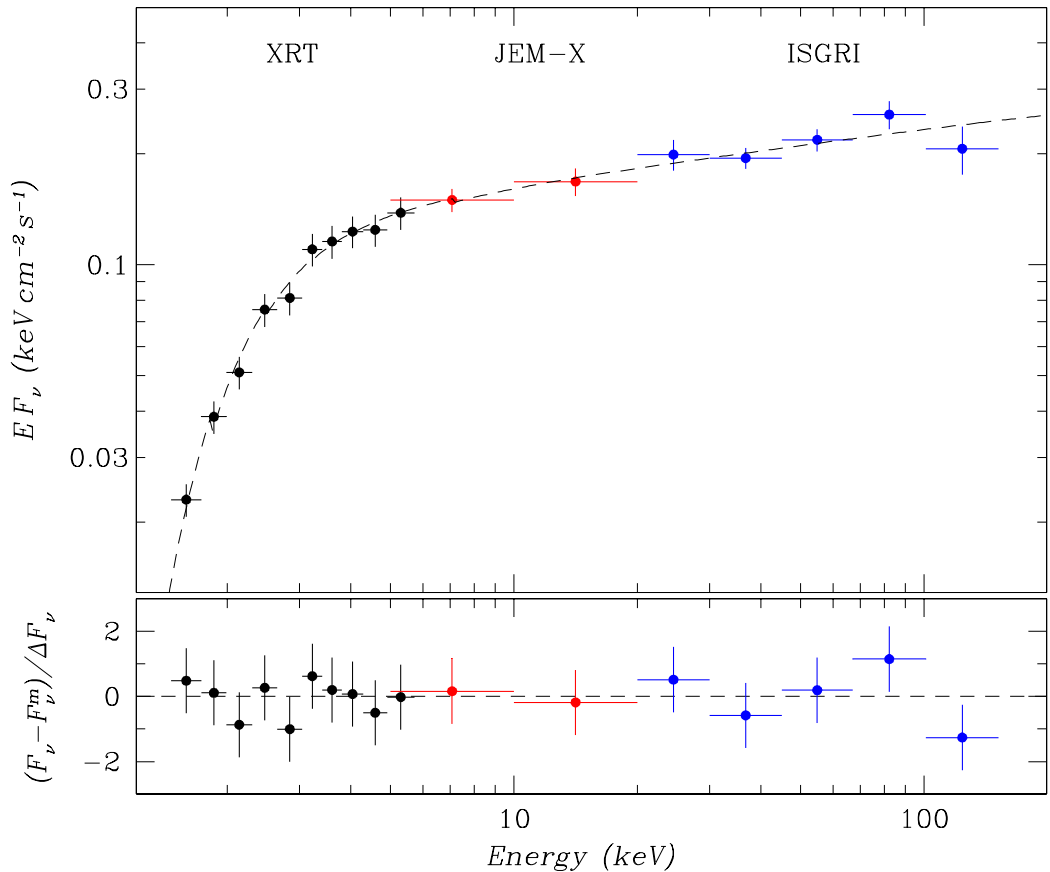}
\vspace{-0.5truecm}
\caption{The broadband X-ray spectrum of \grs. Swift-XRT data are shown in black (ID.~00033812055), JEM-X and IBIS/ISGRI data averaged over 1725-1727 revolutions are shown in red and blue.}
\label{fig:comp_spec}       
\end{figure}

We then constructed broadband X-ray spectra of \grs using quasi-simultaneous observations made by INTEGRAL and Swift. To achieve better statistics for INTEGRAL JEM-X and IBIS/ISGRI data we used the spectrum averaged over 1725-1727 (``plateau'') revolutions. We performed simultaneous fitting of three spectra (XRT 0.3--10 keV, JEM-X 5--20 keV, and IBIS/ISGRI 20--150 keV) using simple {\em const*wabs*powerlaw} model. To account for non-simultaneity of observations and differences in absolute calibrations we introduced cross-calibration coefficients ({\em const}) $C_{XRT}$ and $C_{JEM-X}$), using IBIS/ISGRI data as a reference. The obtained spectrum is presented in Fig.~\ref{fig:comp_spec}. For 1728-1729 revolutions we performed the similar analysis, but without using JEM-X data. 

The best-fit model parameters are presented in Table~\ref{tab:wideband}. The spectrum obtained in the course of ``plateau' ' is well described by a simple model with reasonable $\chi^{2}_{red.}$=0.51 (12 d.o.f.). The surface density of the neutral hydrogen column $N_{H}$ which accounts for the absorption seen in spectra, was measured as $(2.3\pm0.2)\times10^{22}$ cm$^{-2}$ (assuming solar abundance) during the ``plateau'' and  as $(1.7\pm0.2)\times10^{22}$ cm$^{-2}$ at the decline. Both values are close to previously measured $N_{H}\simeq 2.1\times10^{22}$ cm$^{-2}$ \citep{greiner96, miller15_nust}, however they are significantly different from each other. On the other hand, although data in Table~\ref{tab:wideband} show some spectral hardening (from $\Gamma$ = 1.86$\pm$0.07 to $\Gamma$ = 1.73$\pm$0.05) this is not reliable enough. There are no traces of the blackbody radiation or Fe K$\alpha$ fluorescent line. Unlike the {\it NuSTAR} observations from previous outburst \citep{miller15_nust,fuerst16} there is no cut-off at 40--50 keV in the powerlaw component, which we traced up to $\sim$150 keV, thanks to IBIS/ISGRI energy coverage.
We shall note, that observations performed by different telescopes were non-simultaneous and the source showed variability of at least order of 2 during the individual IBIS/ISGRI observations on the ``plateau''. This can explain cross-calibration coefficients of 0.44 and 0.33 for JEM-X and XRT, respectively. This also makes it hard to estimate the true luminosity of \grs. Using IBIS/ISGRI data as reference frame (since it has the largest exposure) we can estimate unabsorbed luminosity in 0.1--100 keV range as L=1.5$\times$10$^{37}$ erg s$^{-1}$, assuming distance to the source of 8.5 kpc, which corresponds to about 1\% of Eddington luminosity for a 10$M_{\odot}$ black hole ( $L_{Edd} = 1.3\times10^{39}$ erg s$^{-1}$).

 \begin{table*}
\caption{The best-fit parameters of broadband X-ray spectra of \grs}              
\label{tab:wideband}       
\centering  
\scriptsize                                    
\begin{tabular}{@{}c c c c c c c c c@{}}    
\hline\hline                        
INTEGRAL     & XRT,    &N$_{H}$,                             & $\Gamma$ ,          & \multicolumn{2}{c}{Flux, 10$^{-10}$, cm$^{-2}$ s$^{-1}$ } &  $C_{JEM-X}$     & $C_{IBIS}$& $\chi^{2}_{red.}$\\    
 revolutions &  ObsID &10$^{22}$ cm$^{-2}$    &   photon index            &        0.1--10 keV               &   20-100 keV                                &                              &                    &  \\ 
\hline
1725-1727&55& 2.33$\pm$0.20                    & 1.86$\pm$0.07  &     8.6$\pm$2.3     &       5.3$\pm$1.4                           &0.44$\pm$0.06&0.32$\pm$0.05 &6.1(12) \\
1728-1729&56& 1.68$\pm$0.19                    & 1.73$\pm$0.05  &     1.8 $\pm$0.1      &     3.0$\pm$0.2                           &--                        &0.56$\pm$0.18 &12.1(10) \\
\hline                                             
\end{tabular}
\vspace{-0.2truecm}
\end{table*}

\begin{figure}
\centering
\includegraphics[scale=0.8]{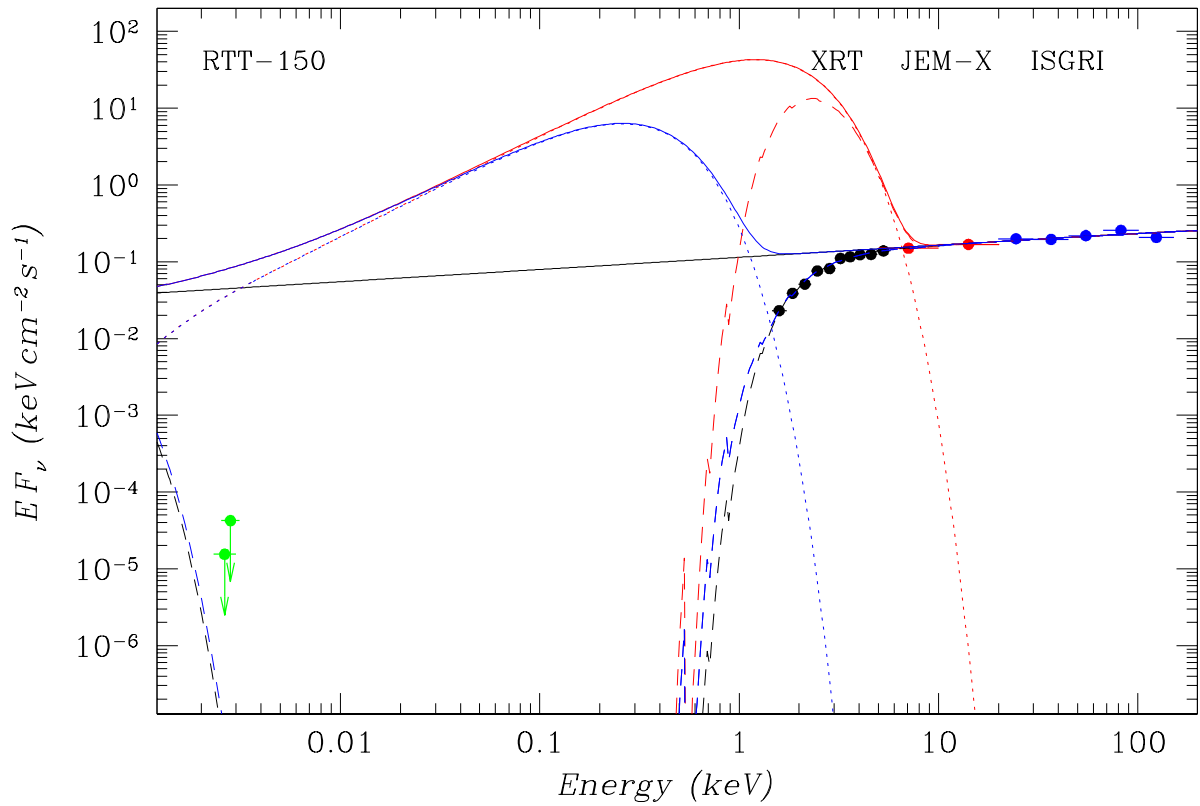}
\vspace{-0.5truecm}
\caption{Same spectra as in Fig.~\ref{fig:comp_spec} but with upper limits from IR observations.  The powerlaw extrapolation of hard X-ray spectra shown in black (solid lines denote true spectrum and dashed one show absorbed). The powerlaw extrapolation with addition of an Eddington-limited cold blackbody disk is shown in red (corresponds to disk with inner radius of 3 $R_{g}$) and in blue (with 30 $R_{g}$). }
\label{fig:wo_spec}       
\end{figure}

No IR source was detected at the position reported by \cite{marti97} in images obtained with RTT-150. We estimated 5$\sigma$ upper limits as $m({r^\prime})\simeq22.0$ and $m({i^\prime})\simeq20.8.$. In Fig.~\ref{fig:wo_spec} we plotted these limits alongside with the spectrum from Fig.~\ref{fig:comp_spec}. Earlier \cite{grebenev13,grebenev14,grebenev16} shown that IR and optical observations of some X-ray novae are well described by the powerlaw extrapolation of their hard X-ray spectra, with respect to the absorption by the dust, corresponding to the photoabsorption by the neutral gas which can be measured from a soft X-ray spectrum. It is interesting to note that observations of X-ray novae in radio bands also suggest single powerlaw \citep{russel06}. From Fig.~\ref{fig:wo_spec} it is seen, that due to severe absorption in direction to the \grs obtained upper limits are much higher than powerlaw extrapolation of the hard X-ray spectrum, as well as possible blackbody contribution from outer and cooler parts of the accretion disk around the black hole. In Fig.~\ref{fig:wo_spec} spectra corresponding to the critical Eddington accretion rate onto 10$M_{\odot}$ black hole are also presented: red lines show spectrum arising from an accretion disk, extended up to the last stable orbit radius (3 R$_{g}$), blue lines show the same disk but with 30 R$_{g}$ inner radius. Both models give equal contribution to the IR band, although the first disk can not be present in this system, because of lack of the soft X-ray emission. We should also note that, under reasonable assumptions, the irradiation can not significantly increase the blackbody disk contribution to the IR emission \citep{grebenev16}.

Based on both spectra and the estimated luminosity we can assume that the source stayed in the canonical low/hard black hole state. Therefore, it turns out that this outburst is failed, at least yet, i.e. amount of accreted matter is not high enough to reach the high/soft state with the prominent blackbody component, that is usual for developed outbursts. Forthcoming  {\it NuSTAR} and {\it Chandra} observations \citep{neilsen16} could shed additional light on this question.
 
 We carried out several addtitonal observations of \grs with IBIS/ISGRI in 1730-1732 revolutions after the paper was accepted. The source remains active with 20--60 keV flux about 15~mCrab. 
 
\section*{Acknowledgments}
This work is based on observations with INTEGRAL, an ESA project with instruments and the science data centre funded by ESA member states (especially the PI countries: Denmark, France, Germany, Italy, Switzerland, Spain), and Poland, and with the participation of Russia and the USA. INTEGRAL  data  used  here  were  obtained  from  the Russian INTEGRAL Science Data Center. Research has made use of SWIFT data obtained through the High Energy Astrophysics Science Archive Research Center Online Service, provided by the NASA/Goddard Space Flight Center. We take the opportunity to thank the TUBITAK National Observatory (Turkey), the Space Research Institute of the Russian Academy of Sciences, and the Kazan Federal University for their support in using RTT-150. Work was financially supported by RNF grant 14-22-00271. Authors are grateful for MPA for the computational support. 
\clearpage	
\label{lastpage}
\bibliographystyle{astron}
\bibliography{author_en.bib}

\end{document}